\documentclass[aps,pre,twocolumn,superscriptaddress,showpacs]{revtex4-2}
\usepackage{graphicx}
\usepackage{amsmath}
\usepackage{natbib}
\usepackage[usenames]{color}
\usepackage{soul}

\usepackage[normalem]{ulem}

\usepackage{amsfonts}
\usepackage{dcolumn}        
\usepackage{amssymb}
\usepackage{bm,amssymb}

\usepackage{bm,fancybox}                	     

\begin{document}

\newcommand{\bea}{\begin{eqnarray}}
\newcommand{\eea}{  \end{eqnarray}}
\newcommand{\bit}{\begin{itemize}}
\newcommand{\eit}{  \end{itemize}}

\newcommand{\be}{\begin{equation}}
\newcommand{\ee}{\end{equation}}
\newcommand{\ra}{\rangle}
\newcommand{\la}{\langle}
\newcommand{\U}{\widetilde{U}}

\def\bra#1{{\langle#1|}}
\def\ket#1{{|#1\rangle}}
\def\bracket#1#2{{\langle#1|#2\rangle}}
\def\inner#1#2{{\langle#1|#2\rangle}}
\def\expect#1{{\langle#1\rangle}}
\def\e{{\rm e}}
\def\proj{{\hat{\cal P}}}
\def\tr{{\rm Tr}}
\def\H{{\hat H}}
\def\Hdag{{\hat H}^\dagger}
\def\Lop{{\cal L}}
\def\Ehat{{\hat E}}
\def\Edag{{\hat E}^\dagger}
\def\Shat{\hat{S}}
\def\Sdag{{\hat S}^\dagger}
\def\Ahat{{\hat A}}
\def\Adag{{\hat A}^\dagger}
\def\U{{\hat U}}
\def\Udag{{\hat U}^\dagger}
\def\Zhat{{\hat Z}}
\def\Phat{{\hat P}}
\def\Op{{\hat O}}
\def\id{{\hat I}}
\def\x{{\hat x}}
\def\P{{\hat P}}
\def\Px{\proj_x}
\def\Pr{\proj_{R}}
\def\Pl{\proj_{L}}


\title{Lagrangian descriptors for the Bunimovich stadium billiard}

\author{Gabriel G. Carlo}
\email[E--mail address: ]{carlo@tandar.cnea.gov.ar}
\affiliation{Comisi\'on Nacional de Energ\'{\i}a At\'omica, CONICET, 
Departamento de F\'{\i}sica, 
Av.~del Libertador 8250, 1429 Buenos Aires, Argentina}

\author{J. Montes}
 \email[E--mail address: ]{jmontes.3@alumni.unav.es}
\affiliation{Departamento de Qu\'{\i}mica, 
 Universidad Aut\'onoma de Madrid,
 Cantoblanco, 28049--Madrid, Spain}
\affiliation{Instituto de Ciencias Matem\'aticas (ICMAT),
 Cantoblanco, 28049--Madrid, Spain}

\author{F. Borondo}
 \email[E--mail address: ]{f.borondo@uam.es}
\affiliation{Departamento de Qu\'{\i}mica, 
 Universidad Aut\'onoma de Madrid,
 Cantoblanco, 28049--Madrid, Spain}
\affiliation{Instituto de Ciencias Matem\'aticas (ICMAT),
 Cantoblanco, 28049--Madrid, Spain}
\date{\today}

\begin{abstract}

We apply the concept of Lagrangian descriptors to the dynamics on the
Bunimovich stadium billiard, a 2D ergodic system with singular families of trajectories, 
namely, the bouncing ball and the whispering gallery orbits. 
They play a central role in structuring the phase space, which is unveiled here 
by means of the Lagrangian descriptors applied to the associated map on the boundary. 
More interestingly, we also consider the open stadium, which in the optical case (Fresnel's laws) 
can be directly related to recent microlaser experiments. 
We find that the structure of the emission profile of these systems can be easily 
described thanks to the open version of the Lagrangian descriptors.  
\end{abstract}
\maketitle

\section{Introduction}
 \label{sec:intro}

The precise characterization of the structure of the phase space of chaotic systems 
is of importance in many areas, and at the same time a non trivial task. 
In particular, the description of the stable and unstable manifolds 
associated to unstable periodic orbits (POs) has recently benefited from 
the introduction of a classical measure, the Lagrangian descriptors (LDs) \cite{LD1}. 
They can be easily evaluated \cite{LD2,LD3,LD10} 
in comparison with other methods like obtaining a Poincar\'e surface of section. 
Their applications range from chemical dynamics \cite{LDW,LD5,LD6,LD7,LD11} 
to abstract mathematical objects like area preserving maps, 
dynamical systems for which discrete LDs have been devised 
and applied to prototypical cases, like that of the Arnold's cat \cite{LD9}.
In this context, chaotic saddles associated to open \cite{Clauss} or scattering \cite{scattering} 
systems have been studied by a redefinition of the LDs for open maps in \cite{openLDs}. 
In fact, eliminating the trajectories going through an 'opening' in phase space (a given finite area 
region \cite{Novaes}) leaves just a repeller, which is a fractal invariant set. 
The redefined measure adapts the LD tool 
in order to reveal the structure embedded in these repellers while keeping its simplicity. 
As a result, the homoclinic tangles associated to POs belonging to the repeller 
can be very efficiently identified. 
This is a difficult task in general, but we have shown how it can be easily done in the case of 
the open tribaker map \cite{openLDs}, where the obtained results can be contrasted with those of
the associated simple symbolic dynamics. 
This allowed to demonstrate once more the usefulness of LDs.

It is of great interest for both theoretical and experimental reasons to extend these 
concepts to realistic situations. 
This can be achieved, for example by considering billiard systems that are used as 
excellent models for resonator cavities in many devices. 
Moreover, it is also important to implement more general ways to make them open, going beyond the 
just purely projective (complete) openings or abstract partially open regions (see for example Ref.~\onlinecite{openLDs}). 
A very relevant example is the escape of trajectories/rays following Fresnel's laws 
\cite{Wiersig}, which leads to many interesting mathematical properties 
and is directly applicable to optical microdisk cavities with deformed boundaries \cite{WiersigR}.

In this paper, we apply LDs to a 2D ergodic billiard -- the paradigmatic 
Bunimovich stadium (BS) --, more specifically to the corresponding map on the boundary using 
Birkhoff coordinates, which captures all the essential features of the dynamics. 
We find that two families of singular orbits, namely bouncing ball orbits (BBOs) 
and whispering gallery orbits (WGOs) provide the foundations of the chaotic structure associated to 
the closed system phase space. 
Moreover, we open the system by letting some trajectories escape according to both, a completely 
projective and also an optical mechanism. 
The latter corresponds to one of the cavities used in very recent microlasers experiments \cite{Bittner}. 
The enhanced vision of POs and their associated homoclinic tangles provided by LDs for open maps allowed 
us to identify which of the shortest ones are responsible for the main properties of the 
emission in experimental situations. 
This throws new light on the future design of these devices, as well as to all kinds of resonator cavities 
(like microwave, for example). 
Also, it reveals which kind of orbits outside of the purely projective repeller could be relevant in 
the semiclassical theory of short POs for partially open systems \cite{STOS1,STOS2}.
 
We have organized this paper in the following way: in Sec.~\ref{sec:LDsforBS} we define the LDs used, 
we describe the main properties of the BS and also those of the corresponding map on the 
boundary written in Birkhoff coordinates. In Sec.~\ref{sec:results} we apply this definition 
to unveil the underlying structure of the main manifolds that characterize the BS phase space in 
the closed case and more importantly, to explain the properties of the emission in the optically 
open scenario. 
Our conclusions are finally presented in Sec.~\ref{sec:conclusions}.

\section{Lagrangian descriptors and the Bunimovich stadium billiard}
 \label{sec:LDsforBS}

Since chaotic maps are prototypical models that capture all the 
essential features of chaotic dynamics, we restrict to them in this work. 
Open maps are obtained by considering a region of escape, in our case 
a two-dimensional 'hole' in our phase space expressed in terms of 
canonical variables ($q$ and $p$). This gives rise to an invariant set, 
the fractal dimensional repeller, given by the 
intersection of the forward and backwards trapped sets. 
These sets are built by computing the trajectories that never escape 
either in the past or in the future, respectively. 
The invariant set can be represented by a measure $\mu(X_i)$ 
which varies in the different regions of phase space $X_i$, 
and it is obtained by the average intensity $I_t$ in the limit $t \rightarrow \infty$, 
considering a large number $N$ of random initial conditions in $X_i$. 
Starting from $I_0=1$, trajectories decrease their intensity according to 
$I_{t+1}=F(q,p) I_t$, which takes into account each hit at the opening.
The function $F(q,p)$ here is arbitrary and characterizes the properties of the 
phase space opening. 
A finite time approximation to this measure (and the repeller) can be obtained as 
$\mu_{t,i}^{b} \cap \mu_{t,i}^{f}$, where 
$\mu_{t,i}^{b,f}=\langle I_{t,i} \rangle/\sum_i \langle I_{t,i} \rangle$ are the finite 
time backwards and forward trapped sets ($\langle \rangle$ stands for the average over 
the initial conditions of each $X_i$).

To understand the structure of the repeller, 
we use in this work the definition of LDs for open maps of 
\cite{openLDs}, adding the contribution of $N$ trajectories $\{q_t,p_t\}_{t=-t_f}^{t=t_f}$ 
at each $X_i$, where $t\in \mathbb{N}$, as  
\begin{equation}
  \begin{array}{lc}
 {\rm LD}={\rm LD^-} \times {\rm LD^+}=\\ 
 \displaystyle\sum_{t=-t_f}^{t=-1} (|q_{t+1}-q_t|^a+|p_{t+1}-p_t|^a) I_t \;\times \\ 
 \displaystyle\sum_{t=0}^{t=t_f-1} (|q_{t+1}-q_t|^a+|p_{t+1}-p_t|^a) I_t, 
   \end{array}
 \label{Eq:LDO}
\end{equation}
and normalizing the LD to $1$.
We use this definition of discrete LDs for open maps, even for the area preserving case, 
since the structure of the manifolds is also better identified in this way. 
We take a $2^8 \times 2^8$ square grid on the torus, $N=10^3$ trajectories at each 
region $X_i$ defined by this grid, and $t_f=30$ (unless otherwise specified). 

\begin{figure}
\includegraphics[width=7.0cm]{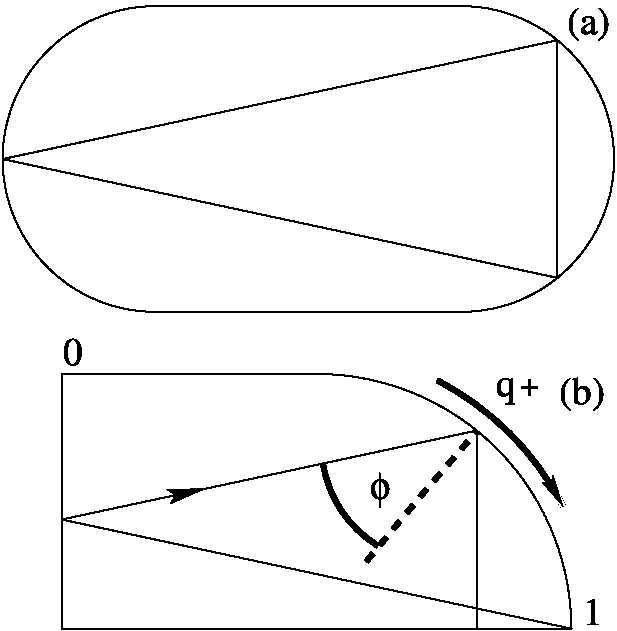}
\caption{(a) Bunimovich stadium billiard with one of the shortest PO as an example. 
(b) Quarter (desymmetrized) billiard with the corresponding version of the PO shown in (a). 
Notice the positive direction of the $q$ coordinate indicated by the curved arrow, 
and the tangential momentum $p=\sin{\phi}$ also positive in this case.
}
\label{fig1}
\end{figure}

The map to which we apply this measure is the surface of section corresponding to the phase space 
of the BS, shown in Fig.~\ref{fig1}(a). 
Notice that any differentiable curve equipped with a coordinate $q$ and 
another $p$ orthogonal to it could have been used alternatively for our purpose. 
When considering billiards that are bounded by rigid walls, 
one usually takes this boundary as the surface of section in Birkhoff coordinates, 
i.e., the boundary \textit{arclength} $q$ and the tangential momentum $p$ at the bounces. 
In fact, we consider just the boundary of the quarter (desymmetrized) billiard measured from 
the upper-left corner and up to the right one (corresponding to $q=0$ 
and $q=1$, respectively), and the tangential momentum $p=\sin{\phi}$, where $\phi$ is the incident 
angle with respect to the normal, as indicated in Fig. \ref{fig1}(b). 
This reduction to the fundamental domain captures 
all the dynamical properties of the full BS, making at the same time a much more efficient 
use of the computational resources by eliminating redundant calculations. 
%

Finally, we define the opening function to obtain the open map on the boundary as the composition 
of the corresponding closed one with the function. 
In this work, we consider two different functions in phase space.
First (for illustrative purposes) $F_R(q,p)$, with $R$ being a reflectivity parameter 
that goes from complete escape for $R=0$ to the closed map when $R=1$ \cite{STOS2}. 
The opening region, in this case, is the domain given by the angle of total
internal reflection which translates into the corresponding $p_{\rm tr}=1/n$ value by using Snell's law, 
where $n$ is the refraction index of the material from which the BS cavity is fabricated. 
As such, this region is delimited by $-p_{\rm tr} \leq p \leq p_{\rm tr}$ for all $q$.
More precisely, we take a constant function given by the value of $R$ in the opening and $1$ elsewhere. 
In this case, we select a complete opening, i.e. $R=0$ (some amount of the incoming trajectories is reflected 
for other values). 
The second reflection mechanism is given by a Fresnel type 
reflectivity function \cite{Wiersig} that involves a partial escape at a dielectric interface. 
For transverse magnetic polarization it is given by
\begin{equation}
F_n(q,p)=\left[ \frac{\sqrt{1-(n p)^2}-n \sqrt{1-p^2}}{\sqrt{1-(n p)^2}+n \sqrt{1-p^2}} \right]^2, 
 \label{reflectivity}
  \end{equation}
the opening region being the same as in the previous case.

\section{Results}
 \label{sec:results}

\begin{figure}
\includegraphics[width=8.5cm]{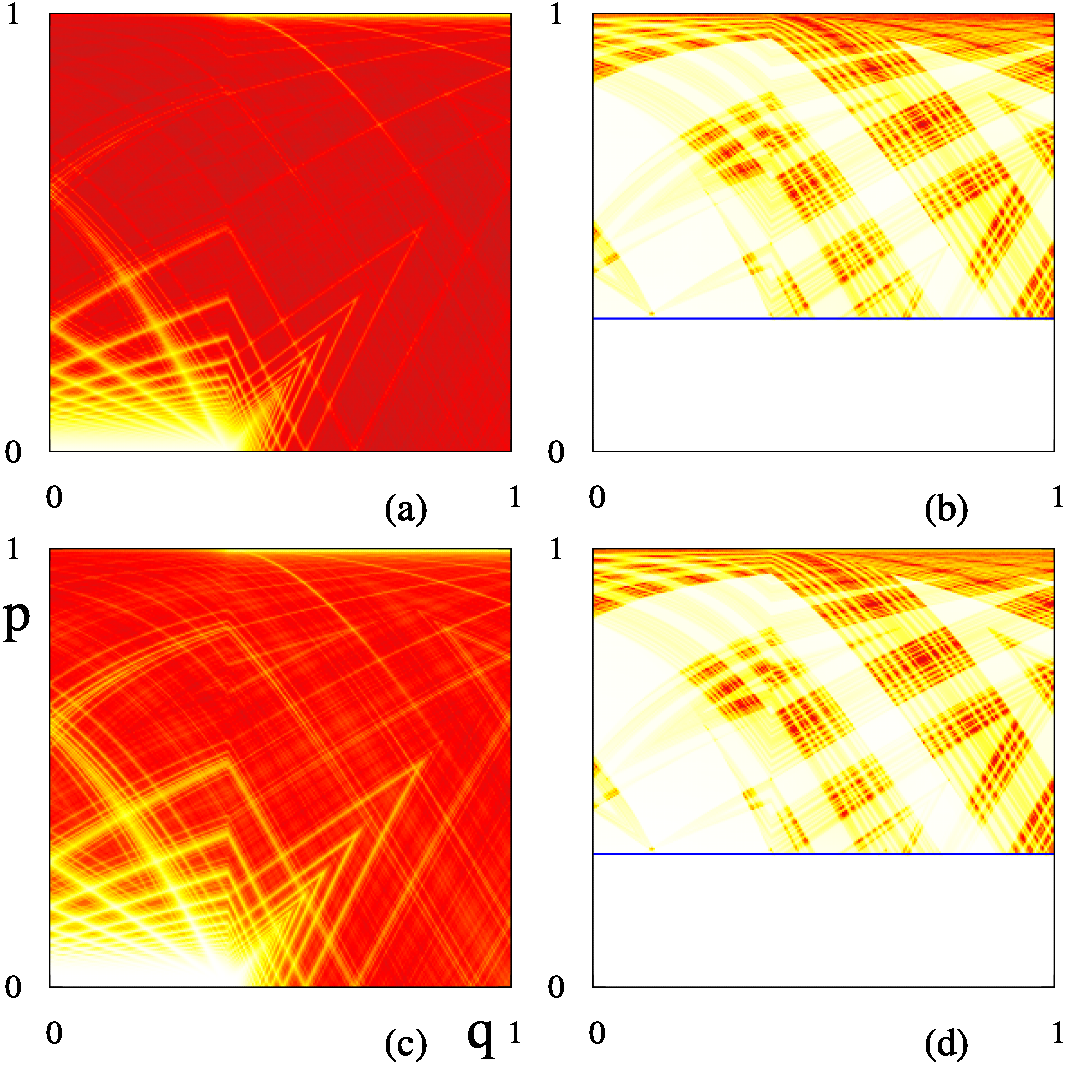}
\caption{(color online) LD for the Bunimovich stadium map on the boundary. 
The left panels correspond to the closed Bunimovich stadium, 
and the right ones show the results for the completely open case with $n=3.3$. 
The upper panels [(a) and (b)] correspond to $a=0.3$ while the lower ones [(c) and (d)] are for $a=0.9$. 
In the right panels the value of $p_{\rm tr}$ is indicated by means of (blue) black horizontal lines.
}
\label{fig2}
\end{figure}

We first apply LDs to study the closed BS. 
The results are presented in Figs.~\ref{fig2}(a) and (c) for two different values of the parameter in 
Eq.~(\ref{Eq:LDO}): $a=0.3$ and $0.9$, respectively. 
We have evaluated the behavior of LD varying the $a$-norm 
in order to check if there is any important dependence on it, when identifying 
the structures in the phase space. 
Our results show that aside of an enhancement of the some structures and a 
blurring of others, no qualitatively different features can be found due to the variation of the used norm. 
Moreover, it is clear that the structure of the chaotic phase space 
is mainly dominated by the manifolds emanating from the BBOs family. 
This is a family of singular, i.e. marginally stable trajectories, 
that constitute a continuous set along the lower values of the boundary coordinate (corresponding 
to the straight segment of the wall), impacting at an angle $\phi=0$. 
Also, the WGOs that develop mainly along the boundary with very large values of $\phi$, 
play an important part in the phase space structure. 
Indeed, they give rise to the next relevant sets of manifolds emanating from them, which are 
essentially located on a very narrow region corresponding to the circular part of the boundary near $p=1$. 
Next, in Figs.~\ref{fig2}(b) and (d) we present the effect on this set 
of considering a complete opening in the BS for $n=3.3$. 
As can be seen, the BBOs are completely wiped out along with their manifolds, 
while the WGOs survive in the repeller. 
Actually, the latter are the dominating POs (with their manifolds) for this case, 
completely ruling the structure of the phase space. 
%

\begin{figure}
\includegraphics[width=8.5cm]{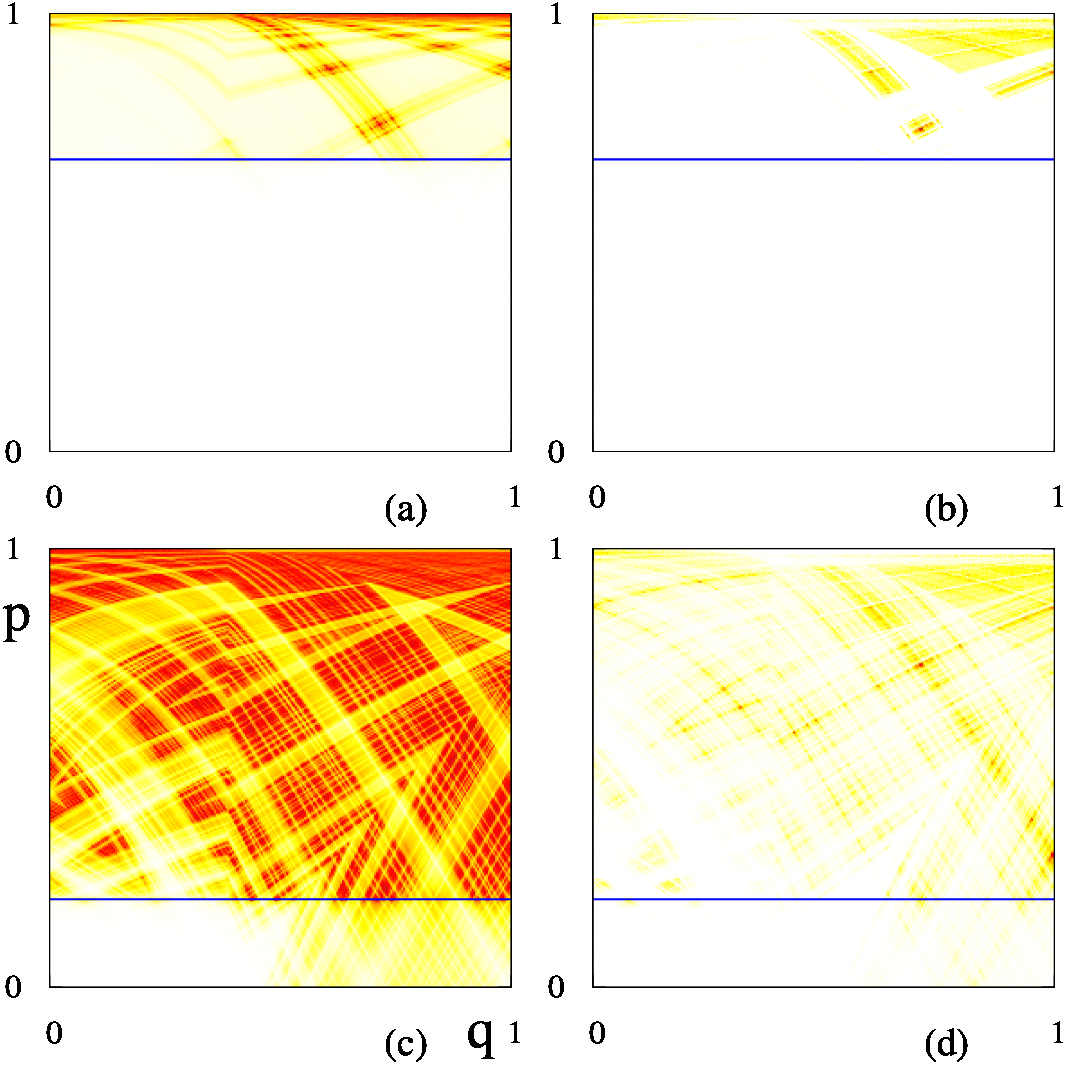}
\caption{(color online) LD for the partially (Fresnel) open Bunimovich stadium map on the boundary. 
The left panels correspond to $a=0.3$ while the right ones to $a=-0.3$. 
The upper panels show the $n=1.5$ case and for the lower ones $n=5.0$. 
In all panels the value of $p_{\rm tr}$ is indicated by means of (blue) black horizontal lines. 
}
\label{fig3}
\end{figure}

We now further analyze the partially (Fresnel) open BS in order 
to characterize the structure of the repeller beyond the clear dominance of the WGOs. 
The top and bottom panels of Fig.~\ref{fig3} show the results for two extreme cases, 
corresponding to low and high values of $n$, respectively. 
In Fig. \ref{fig3}(a) we show the case $n=1.5$ ($a=0.3$) where the $p_{\rm tr}$ value is high. 
There is a big opening through which most of the trajectories escape and the only relevant surviving 
ones are the WGOs with a few more orbits contributing to the structure of the repeller. 
To find out which are the most significant, we take $a=-0.3$ and eliminate 
the WGOs [see results in Fig.~\ref{fig3}(b)]. 
This leaves just one relevant small area that is far from the $p_{\rm tr}$ line. 
On the other hand, at the other extreme, i.e. $n=5.0$, 
we see that though WGOs are again the main component of the repeller [see Fig. \ref{fig3}(c) 
where we used $a=0.3$], there are many more POs and associated manifolds building it.
This is due to the fact that the opening is much smaller than in the previous case. 
Also, there is a significant portion of the repeller in the opening, rendering a morphology which is
different from that of the completely open or even the previous case.
Again, by taking $a=-0.3$ and eliminating the WGOs, we can see that just two small regions 
mainly build the repeller, and that they are far from the $p_{\rm tr}$ line 
[see Fig.~\ref{fig3}(d)].
%

\begin{figure}
\includegraphics[width=6.5cm]{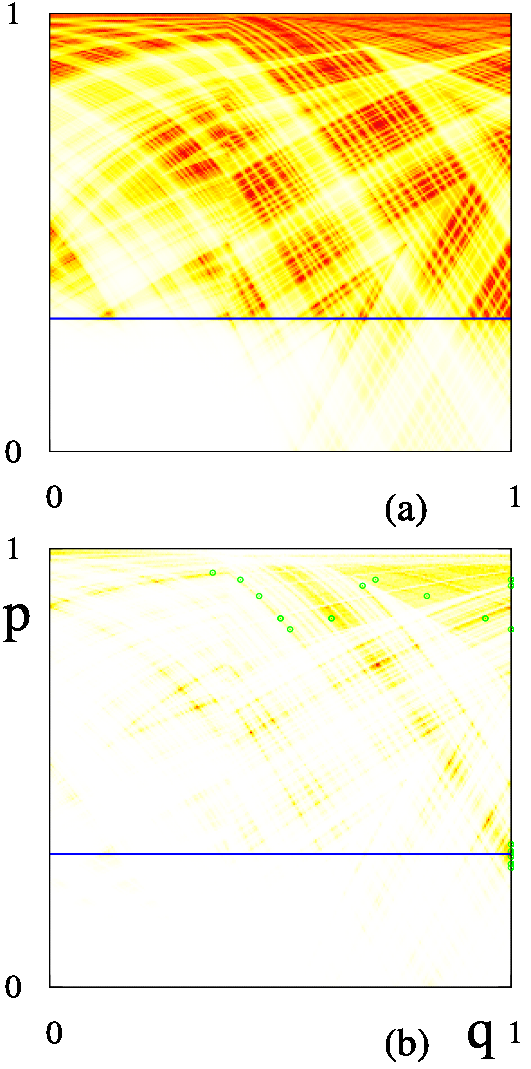}
\caption{(color online) LD for the partially open Bunimovich stadium map on the boundary, 
following Fresnel's laws with 
$n=3.3$. The value of $p_{\rm tr}$ is indicated by means of (blue) black horizontal lines. 
In (a) $a=0.3$ and in (b) $a=-0.3$. The (green) gray circles in (b) show 
all the periodic points associated to the shortest POs that pass through the enhanced region around 
$p_{\rm tr}$.
}
\label{fig4}
\end{figure}

In a recent series of very interesting experiments \cite{Bittner} the emission at the boundary of 
a partially open BS following Fresnel laws has been studied both theoretically and experimentally.
In order to compare with these experimental work, we take $n=3.3$ (considered in \cite{Wiersig}), 
which is approximately the experimental value and evaluate the same measures as before. 
In Fig.~\ref{fig4}(a) we take $a=0.3$ to  obtain the corresponding LD, 
which permits to see the WGOs, as well as more POs and manifolds building 
the repeller (less than in the Fig.~\ref{fig3}(c) case though). 
As can be seen, there is also a portion of it lying inside the opening. 
In order to reveal its structure beyond the influence of the WGOs, we again take $a=-0.3$.
The corresponding results are shown in 
Fig.~\ref{fig4}(b), where two small regions can be noticed as the major 
contribution to the repeller structure. 
One of them is located at both sides of the $p_{\rm tr}$ line (at around $q=1$). 
To better understand the physical meaning of this region, 
we have conducted a systematic search through the first 200 POs of the BS reaching a maximum number 
of bounces $T=14$ with the full boundary [see Fig. \ref{fig1}(a)]. 
We have checked which of them pass through this enhanced region. 
The results of this search are shown using (green) gray circles 
in Fig.~\ref{fig4}(b). 
These are all POs that have periodic points in the map on the boundary that are either in the neighborhood
of WGOs or around the $p_{\rm tr}$ line.
%

\begin{figure}
\includegraphics[width=7.5cm]{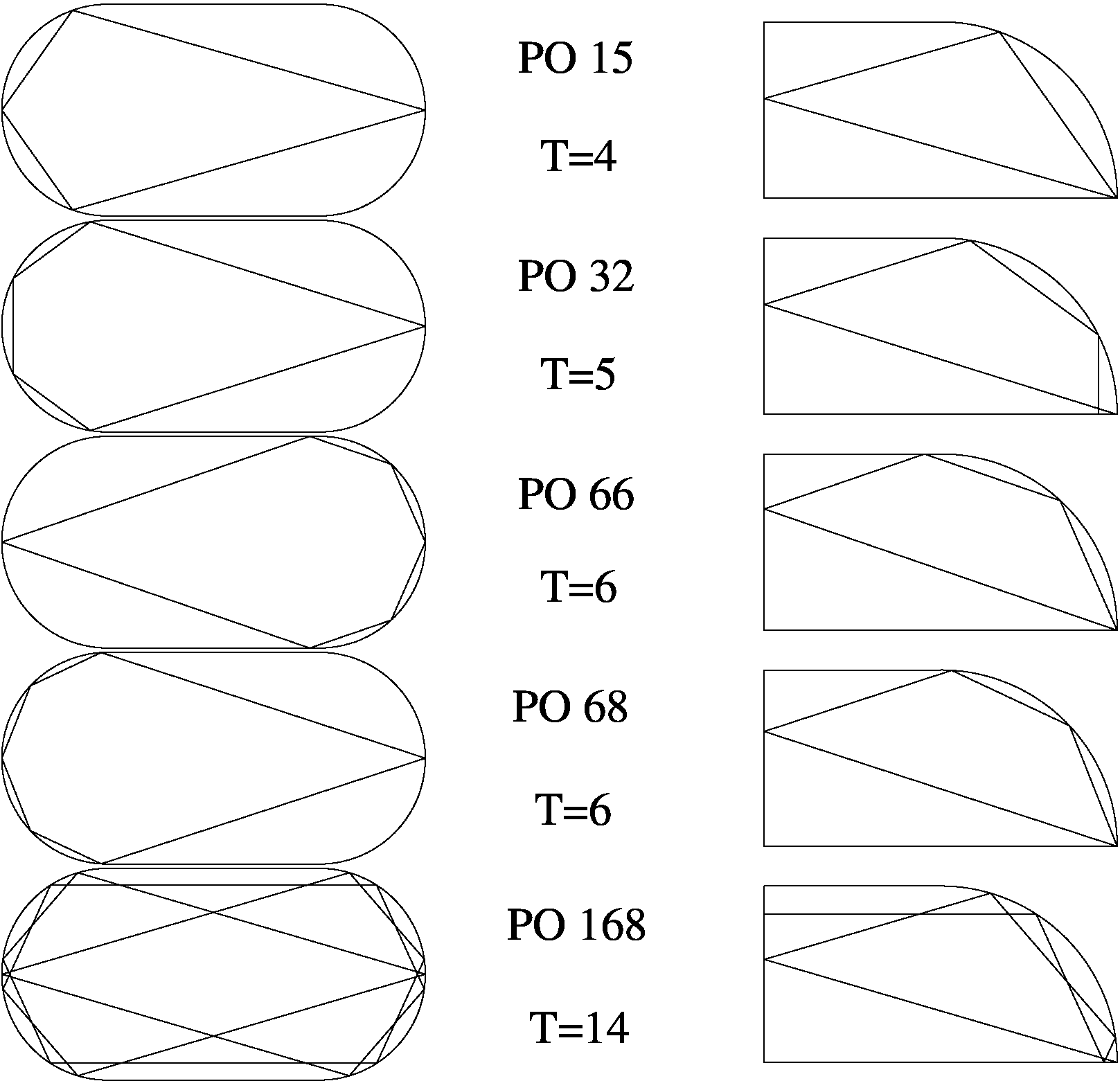}
\caption{Shortest POs that pass through the enhanced region of LD around $p_{\rm tr}$ 
shown in Fig.~\ref{fig4}(b). In the left and right columns the full and desymmetrized Bunimovich 
stadium versions are shown. The POs are ordered according to their lengths, and the number of bounces 
$T$ for the full case is indicated for each one of them. 
Notice that PO $168$ is the only fully symmetrical (with respect to horizontal and vertical reflections), 
we show an example of a symmetry related partner by means of PO $66$.
}
\label{fig5}
\end{figure}

How do these relevant POs look like? 
To gain some insight about their morphology, we present them in the left column of Fig.~\ref{fig5}, 
plotted in configuration space, by ascending length order (from top to bottom). 
The first four are some of the shortest possible POs, and only the last one is a bit longer. 
Nevertheless, they all share a common main property: they have a large portion that resembles the WGOs, 
and make one or two bounces near $p_{\rm tr}$. 
In fact, they connect the highly confined regions of the repeller with the opening. 
The desymmetrized version of these POs, shown in the right column, are illustrative of the fact that the
economy of computational resources provided by this approach does not affect capturing all features of the dynamics.
%

\begin{figure}
\includegraphics[width=8.0cm]{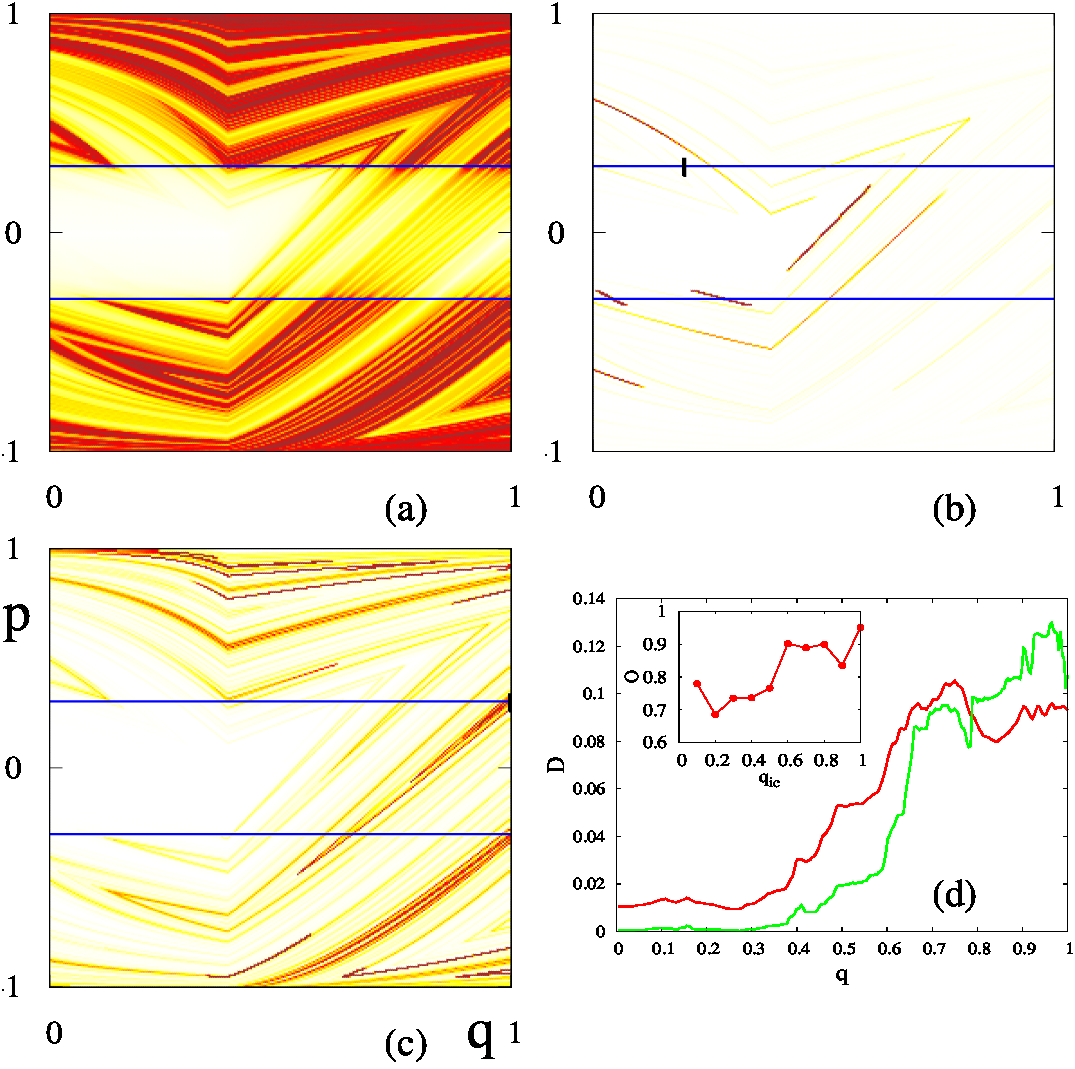}
\caption{(color online) (a) ${\rm LD}^+$ for the partially open Bunimovich stadium map on the boundary, 
following Fresnel's laws with $n=3.3$, which corresponds to the unstable manifold. 
In (b) and (c), L distributions of $10^6$ random initial 
conditions located at the black rectangles around $p_{\rm tr}$ ($q_{\rm ic}=0.2$ and $q_{\rm ic}=1.0$, respectively) 
and evolved up to time $t_f=14$ are shown. In all cases, the values of $\pm p_{\rm tr}$ are indicated by 
means of (blue) black horizontal lines.(d) Normalized distribution D at the opening region (from $-p_{\rm tr}$ to 
$p_{\rm tr}$) as a function of $q$ for the ${\rm LD}^+$ in (a) ((red) black line) and for the distribution in (c) 
((green) gray line). In the inset we show the overlap O (restricted to the opening) 
between the ${\rm LD}^+$ and similar distributions to those in (b) and (c) but for different values of $q_{\rm ic}$. 
}
\label{fig6}
\end{figure}

Finally, we address an important related point, namely which is the role played by these POs 
(and its neighboring region/homoclinic tangle) in the microlaser emission? 
To answer this question, we have calculated the unstable manifold of the partially open BS 
using the ${\rm LD}^+$ measure; the results are shown in Fig.~\ref{fig6}(a). 
Besides the fact that most of the escape from the repeller happens at the 
circular part of the boundary, it is hard to see any finer structure associated to a given value of $q$. 
Moreover, it is worth noticing that determining the structure of this manifold is not always straightforward 
\cite{Bittner,microlasers2}, however the LDs used here allowed us to do that in an extremely simple way. 
Also recall that the unstable manifold method \cite{Noeckel1,Noeckel2,Noeckel3} it is widely accepted 
in order to make a theoretical prediction of the microlasers emission patterns \cite{microlasers2}, 
at least when the wavelength is much smaller than the cavity size, i.e. deep in the semiclassical regime. 
In this respect, some works focused on the inner structure of the emission, pointing towards POs 
of interest. In \cite{Schwefel} the directionality of the far field emission was related to very short POs, 
in particular the rectangle orbit of a specific case of the BS with a short straight segment (thought as a deformation of boundaries associated to mixed phase spaces). 
Also, in \cite{microlasers3} a stable rectangle PO plays a central 
role in a mixed dynamical scenario (and in the quantum regime). 
We next make systematic our analysis by calculating the short time forward evolution ${\rm L(}q,p)$ for $10^6$ random initial conditions, located in small domains around $p_{\rm tr}$ in phase space. 
We consider rectangles of a size given by one region $X_i$ along $q$ and nine along $p$, 
centered at $p_{\rm tr}$ and at $q_{\rm ic}=0.1,\ldots,1.0$ with $\Delta q=0.1$. 
The evolution is then carried out up to $t_f=14$, which is the maximum number of trajectory bounces 
allowed in our previous search. 
The results are shown in Figs.~\ref{fig6}(b) and (c), where we can see two significant examples of this. 
In the first case $q_{\rm ic}=0.2$ was considered for the initial distribution, 
while in the second $q_{\rm ic}=1.0$ (see the small rectangles around $p_{\rm tr}$). 
It is clear that for the smaller $q_{\rm ic}$ value the evolved density does not reveal the features of 
the emission region of the repeller, while it does for $q_{\rm ic}=1.0$.
Obviously, some differences are observed between the two cases, 
and we do not claim that just a single orbit is responsible for all of the emission, but we do think that 
a small set of them associated to an homoclinic tangle can provide a very complete approximation to its structure. 
As a matter of fact, if we look at results in Fig~\ref{fig6}(d), it can be seen that the normalized 
cumulative distribution in the opening ${\rm D}=\int_{-p_{\rm tr}}^{p_{\rm tr}} {{\rm L}} \: dp$ as a function of $q$
resembles that of the complete unstable manifold ${\rm D}=\int_{-p_{\rm tr}}^{p_{\rm tr}} {{\rm LD}^+} \: dp$. 
In the inset we see that the overlap 
${\rm O}=\int_{-p_{\rm tr}}^{p_{\rm tr}} \int_{q} {{\rm LD}^+ \times {\rm L}} \: dq dp$ 
between the full distributions in the escaping region corresponding 
to the unstable manifold and those at different $q_{\rm ic}$ are the worst for $q_{\rm ic}=0.2$ and the best 
for $q_{\rm ic}=1.0$, reaching here a remarkable $0.95$ value. 
\section{Conclusions}
 \label{sec:conclusions}

In this work we have extended the application of the LD classical measure to realistic open systems. 
By considering the partially open BS with Fresnel's laws we have addressed an issue of great experimental 
interest, such as the emission properties of microlasers and dielectric resonant cavities in general 
\cite{Bittner,microlasers3}.
Nevertheless the LDs of the closed system revealed very useful in understanding the central role played 
by BBOs, followed in relevance by the WGOs and their associated manifolds, in structuring the chaotic phase space. 
Opening the system induces the formation of a repeller whose main characteristic is the lack of the BBOs 
and their main manifolds, whose role is now taken over by the WGOs. 

By using our definition of LDs for open systems \cite{openLDs} we were able to unveil the inner structure of the 
repeller, specially in the escaping region. 
We have been able to identify a set of short POs that pass through a small enhanced 
region of the repeller that is located near $q=1.0$ and $p=p_{\rm tr}$. 
We have found that they share the same morphology, being a kind of hybrid between WGOs 
(that act as a sort of reservoir) and escaping trajectories. 
Given that the partially open BS microlaser emission on the boundary \cite{Bittner} can be determined 
by means of the unstable manifold calculation, we have computed this magnitude by using our measure 
in a very straightforward way. 
Moreover, we compared this distribution with the ones obtained by means of a short time evolution 
of initial conditions very close to the POs identified thanks to the LDs. 
We have been able to quantitatively demonstrate in this way that these POs provide all the relevant 
structure of the escaping portion of the repeller. 
This leads us to conclude that the interplay between the shape of the microcavity and the index of refraction 
$n$ can single out a small group of short POs than can very well approximate the emission properties 
in chaotic systems. Our method provides a systematic way to find them. This is a very important result 
that should have a large influence in the future design of microlasers \cite{Li,Andreoli}. 

Finally, we think that this work could be of relevance at the time of 
considering smaller cavities, where a purely classical calculation would not be enough to describe them. 
In that scenario, the semiclassical theory of short POs \cite{STOS1,STOS2} will benefit from the identification 
of those orbits outside of the completely open repeller that significantly contribute to construct the 
partially open one in realistic situations.

\section{Acknowledgments}
Support from CONICET is gratefully acknowledged.
This research has also been partially funded by the Spanish Ministry of Science, 
Innovation and Universities, Gobierno de Espa\~na under Contract No.~PGC2018-093854-B-I00,
and by ICMAT Severo Ochoa Programme for Centres of Excellence in R \& D
(CEX2019-000904-S).

%
\end{document}